\documentstyle[12pt]{article}

\newcommand{\be}{\begin{equation}}
\newcommand{\bea}{\begin{eqnarray}}
\newcommand{\ee}{\end{equation}}
\newcommand{\eea}{\end{eqnarray}}
\newcommand{\partud}{
{d \sigma (\uparrow\uparrow-\uparrow\downarrow) \over dx dCos(\theta)}}

\newcommand{\partg}{ {
d \sigma^g(s,s_1) \over dz d Cos(\theta)}}

\newcommand{\partG}{
{d \sigma^g (\uparrow\uparrow-\uparrow\downarrow) \over dz dCos(\theta)}}

\begin{document}
\begin{titlepage}
\vspace*{\fill}
\begin{center}
{\Large \bf Gluon fragmentation function in polarised\\$\Lambda$ hyperon
production:
The Method of Factorisation 
}\\[1cm]
{\bf V. Ravindran}\\
{\em Theory Group, Physical Research Laboratory, Navrangpura \\
Ahmedabad 380 009, India}\\
\end{center}
\vspace{2cm}
\begin{abstract}
We discuss the polarised fragmentation functions of quarks and
gluons in Perturbative Quantum Chromodynamics.  The Altarelli-Parisi
evolution equations governing these fragmentation functions
are presented.  We find that the first moment of the polarised
gluon fragmentation function to order $\alpha_s$ is as important
as that of the quark and anti-quark fragmentation functions.
The $\Lambda$ production in polarised $e^+ e^-$ annihilation where
this can be realised is discussed.   We propose the
factorisation method to compute the gluonic contribution
to $\Lambda$ production to order $\alpha_s$.  
With appropriate operator definitions for the quark and
the gluon fragmentation functions, we find that the hard part of the
factorisation formula is free of any infrared singularities 
confirming the factorisation.  
 
\end{abstract}
\vspace*{\fill}
\end{titlepage}
Quantum Chromodynamics(QCD) is the most successful theory of 
strong interaction physics describing the dynamics of the 
quarks and gluons inside the hadrons $\cite {MUTA}$.  
Parton Model description of various strong interaction processes
has been a successful description.
Both inclusive processes such as Deep Inelastic Scattering(DIS) and 
hadro production in   
$e^+ e^- $ annihilation can well be described
in the QCD improved Parton model.
In DIS, the probability distribution functions 
of quarks and gluons inside the incoming hadrons are the non-perturbative
inputs of the model.  Though they are non-perturbative, 
their evolution in terms of the scale
is completely described by the well known Altarelli-Parisi(AP) evolution
equations.  In the processes of hadroproduction in $e^+ e^-$ annihilation, 
the fragmentation functions of quarks and gluons into hadrons are 
the non-perturbative inputs but again their evolution is completely 
governed by the AP evolution equations.
%Scaling violations in DIS is well understood 
%so as that in the processes of hadroproduction $\cite {NASON}$.  

Needless to say that 
Perturbative QCD(PQCD) improved Parton model unravelled 
both momentum and spin structure of 
the proton in terms of its constituents such as quarks and gluons.
The momentum structure of the proton was studied in the unpolarised 
lepton proton DIS experiments $\cite {MUTA}$.  Few years 
ago a series of polarised 
lepton proton DIS experiments unravelled the contributions coming 
from quarks and gluons to the proton spin $\cite {ANS}$.  
In both the unpolarised and the
polarised DIS experiments, a remarkable fact emerged which was that the 
gluons share significant amount of the momentum and the spin of the proton 
contradicting the naive expectation that most of the 
momentum and the spin are carried by the valence quarks.  These 
experimental facts are well understood by using the AP
evolution equations of both unpolarised and polarised quark
and gluon distribution functions.  These inclusive experiments are excellent
tests of QCD improved Parton model and the AP evolution equations.

Recently, there has been a lot of works
on the fragmentation functions of quarks and gluons measurable
in the process of hadroproduction in $e^+ e^-$ annihilation and 
also on the test
of QCD improved Parton model sublimented with the AP evolution equations for
the fragmentation functions.  All these analysises are on unpolarised
$e^+ e^- $ scattering experiments which require only the AP equations
for unpolarised fragmentation functions.   The QCD improved analysis of the
polarised $e^+ e^-$ scattering experiment is yet to be done.  

It is in this 
spirit, this letter presents a systematic analysis of the polarised 
fragmentation functions of quarks and gluons within PQCD.  
This analysis is important as we will see
that polarised gluonic contribution to the production of polarised
hadrons is as important as polarised quark contribution to polarised
hadrons.  Our analysis is quite general in the sense that it can be 
applied to other inclusive hadroproduction processes such as polarised 
proton proton and polarised proton anti-proton scattering experiments 
as these fragmentation functions and their evolutions
are universal.  

Let $D_{a(h)}^{H(s)}(x,Q^2)$ be the probability 
distribution for a parton $a$ with polarisation $h$ produced 
at a scale $Q^2$ to fragment into
a hadron($H$) of polarisation $s$ carrying a momentum fraction $x$ of 
the quark momentum.  Now we define polarised parton
fragmentation functions as
\be
\Delta D_a^H(x,Q^2)= D_{a(\uparrow)}^{H(\uparrow)}(x,Q^2)- 
D_{a(\uparrow)}^{H(\downarrow)}(x,Q^2)
\ee
If the strong interaction world consists only of quarks and a single hadron,
then the fragmentation function is $\delta(1-x)$ implying that all the spin 
will be transferred to the produced hadron.  This is not usually the 
case because of two reasons.  Firstly the spin of the parent parton 
can be distributed as it can undergo collisions which can flip 
the spin(mass corrections).  
Secondly, the single quark can fragment into many hadrons as it can 
excite the vacuum producing $q \bar q$ pairs.  So the quark spin will be
distributed among the produced hadrons.  So it is not necessary
that the quark spin is completely transferred to a single hadron.
The situation is similar to DIS where the proton spin is
shared by various partons.  

The evolution of the quark and gluon fragmentation functions are
governed by the AP evolution equations and are given by
\be
{d \over dt} \Delta D_{q_i}^H(x,t) = {\alpha_s(t) \over 2 \pi}
\int_x^1 {dy \over y} \left[ \Delta D_{q_i}^H(y,t) \Delta P_{qq}(x/y)
+ \Delta D_g^H(y,t) \Delta P_{gq}(x/y)\right]
\label{apeqnq}
\ee
\be
{d \over dt} \Delta D_g^H(x,t)\! =\! {\alpha_s(t) \over 2 \pi}
\!\int_x^1\!\! {dy \over y} \left[ \sum_{j=1}^{2f}\!\Delta D_{q_i}^H(y,t) 
\Delta P_{qg}(x/y)\!\! +\!\! \Delta D_g^H(y,t) \Delta P_{gg}(x/y)\right]
\label{apeqn}
\ee
where $t=\log(Q^2/\Lambda^2)$ and $\alpha_s(t)$ is the strong coupling
constant.  Here, $\alpha_s(t) \Delta P_{ab}(y) dt$ 
is the probability density of finding
a parton of type $a$ at the scale $t+dt$ with momentum fraction
$y$ inside the parton of type $b$ at a scale $t$.  
The splitting functions are given by
\bea
\Delta P_{qq}(z)&=&C_2(R)\left({1+z^2 \over (1-z)_+}+ {3 \over 2} \delta(1-z) 
\right) 
\nonumber\\
\Delta P_{gq}(z)&=&C_2(R) \left( {1- (1-z)^2 \over z}\right)
\nonumber\\
\Delta P_{qg}(z)&=&{1\over 2} (z^2-(1-z)^2)
\nonumber \\
\Delta P_{gg}(z)&=&C_2(G)\left( (1+z^4) \left({1\over z} + {1\over (1-z)_+}
\right) \right. \nonumber \\
&&\left. -{(1-z)^3 \over z} + \left( {11 \over 6} - {2 \over 3} 
{T(R) \over C_2(G)}\right) \delta(1-z) \right)
\nonumber
\eea
Here, $C_2(R)= (N^2-1)/2N$, $C_2(G)=N$ and $T(R)=f/2$ with $N=3$ for 
$SU(3)$ and $f$ being the number of flavours $\cite {MUTA}$.  In 
the above equations,
usual $+$ prescription has been used.  Notice that the splitting function 
matrix appearing in this AP evolution equation is 
just transpose of that appearing in the AP evolution equation 
for the parton probability distributions.  This can be understood very easily:  
The emission of a quark(gluon) from a quark
(gluon) only affects the probability of quark (gluon) fragmenting 
into hadron.
On the other hand, the emission of a quark from a gluon 
changes the probability of the gluon fragmenting into hadron.  
Similarly, the emission of a gluon from a quark affects the probability
of quark fragmenting into hadron.  That is
why this matrix is transpose.  These equations can easily be solved in
the Mellin space.  Define
\bea
\Delta D_a^H(n,t)&=&\int_0^1 x^{n-1} \Delta D_a^H(x,t) dx
\nonumber \\
\Delta P_{ab}(n)&=&\int_0^1 x^{n-1} \Delta P_{ab}(x) dx
\eea
%In terms of the above definitions,
%be
%d\over dt}{\pmatrix {\Delta D_{q_i}^H(n,t)\cr \Delta D_g^H(n,t)\cr}}
%= {\alpha_s(t) \over 2 \pi} {\pmatrix {\Delta P_{qq}(n) & 
%  \Delta P_{gq}(n)\cr \Delta P_{qg}(n) & 
%  \Delta P_{gg}(n)\cr}} \ocross {\pmatrix {\Delta D_{q_i}^H(n,t)\cr 
%  \Delta D_g^H(n,t)\cr}}
%ee
It is interesting to consider the lowest moment $n=1$.  From 
eqn.(\ref {apeqn}), we find that
the first moment of the polarised gluon fragmentation function to 
order $\alpha_s(t)$ satisfies a simple first order differential 
equation, that is
\be
{ d \over dt} \Delta D_g^H(t) = \alpha_s(t) \beta_0 \Delta D_g^H(t) 
\label{apg}
\ee
where $\beta_0=(11 C_2(G)- 4 T(R))/12 \pi$.
The solution to the above equation can be found very easily 
using renormalisation group(RG) equation for the QCD coupling constant,  
\be
{d \over dt}\alpha_s(t)= - \beta_0 \alpha_s^2(t)
\label{alrg}
\ee
From eqns.(\ref {apg}, \ref {alrg}),
we obtain an interesting behaviour of first moment of gluon 
fragmentation function:  the product of the first moment of
polarised gluon fragmentation function times the strong coupling constant
is scale independent to order $\alpha_s^2(t)$,
\be
{d \over dt} (\alpha_s(t) \Delta D_g^H(t))= 0 (\alpha_s(t)^3)
\label{aldg}
\ee
In other words, to order $\alpha_s^2(t)$, $\Delta D_g^H$ increases 
as the scale $t$ increases, i.e
\be
\Delta D_g^H(t) = K \log \left({Q^2 \over \Lambda^2}\right)
\ee
where $K$ is some constant.  Recall that the 
counter part of such a relation in the polarised parton 
distribution functions exists and has opened up a better understanding of 
the spin structure of the proton $\cite {ANS}$.  That is, due 
to the similar relation 
\be
{d \over dt} (\alpha_s(t) \Delta g(t))= 0 (\alpha_s(t)^2)
\ee
where $\Delta g(t)$ is the first moment of polarised gluon 
distribution function, the polarised gluonic contribution to
proton spin is significant at very high energies.  This was experimentally
observed and the result was that the gluon carries significant amount
of the spin of the proton.  Hence the relation(eqn.(\ref {aldg}))
suggests that at very high
energies the polarised gluon fragmentation into polarised
hadrons is as significant as polarised quark fragmentation into polarised
hadrons.  So care should be taken when one interprets the
data on polarised hadron production in 
polarised $e^+ e^-$ annihilation
processes.  

Now let us consider the first moment of 
the polarised quark fragmentation function into polarised hadron.  
From eqn.(\ref {apeqnq}), it turns out 
\be
{ d \over dt} \Delta D_q^H(t)={1 \over \pi} \alpha_s(t) \Delta D_g^H(t)
\label{aldq}
\ee
From eqns. (\ref {aldg}) and (\ref {aldq}), we find that 
$\Delta D_{q_i}^H(t)$ grows as $t$.  
So this is a remarkable behaviour.
This is very different from first moment of the polarised quarks 
in polarised hadron where it is scale independent.  

It is instructive to find out the consequences 
of these results from theoretical 
point of view.  Let us consider the polarised inclusive 
hadroproduction of $e^+ e^-$ 
scattering. In the quark sector, the first moment of the cross section 
is proportional to 
$\Delta D_q^H(t)$.  The eqn. (\ref {aldq}) suggests that 
the leading behaviour of 
$\Delta D_q^H(t)$ is logarithmic.  In the gluonic sector, the first
moment of the gluonic 
contribution is proportional to $\Delta D_g^H(t)$.  The coefficient 
function is proportional to $log(Q^2)$ times strong coupling 
constant $\alpha_s(t)$.  So, from the eqn. (\ref {aldg}), the first 
moment of the gluonic 
contribution is of the same order as that of leading order quark contribution. 
Hence the naive expectation that polarised gluonic contribution to
polarised hadroproduction is next to leading order effect is not
correct.  The above observation is very general as it is
based on the AP evolution equation for polarised parton fragmentation functions.
So the above analysis is applicable to other polarised hadroproduction
processes.

In the following we consider the polarised hadroproduction in polarised
$e^+ e^-$ annihilation $\cite {BUR}$.  Since the gluonic contribution 
is as important
as quark contribution, we compute the polarised gluonic contribution
to polarised $\Lambda$ production.
We work in the energy range where only process
where 'photon' fragmenting into hadron is dominant.  
The reason why we consider inclusive polarised $\Lambda$ production is that
the cross section is large compared to that of other particles.
Also it is easy to measure the polarisation of $\Lambda$ $\cite {LUN}$.  
Our analysis can be extended to other
hadro productions. The complete analysis 
including $Z$ exchange channel is reserved for future publication 
$\cite {RAV1}$.

The inclusive polarised $\Lambda$ production rate factorises as:
\be
d \sigma(s,s_1) = {1 \over 4 q_1.q_2} L^{\mu \nu} 
(q_1,q_2,s_1) ( {e^2 \over Q^4}) 4 \pi W_{\mu \nu}^\Lambda (q,p,s)
{d^3 p \over (2 \pi)^3 2 p_0}
\ee
where $L_{\mu \nu}(q_1,q_2,s_1)$ is the leptonic part arising from $e^+ e^-$
annihilation into a photon of virtuality $Q^2$ and
$W_{\mu \nu}^\Lambda(p,q,s)$ is photon fragmentation tensor.
The arguments of these tensors are momenta
described in the figures 1 and 2 and $s,s_1$ are the spins of $\Lambda$ 
and electron respectively.  The photon fragmentation tensor 
contains all the information about 
the polarised quark and gluon fragmentation into polarised $\Lambda$.  

The operator definition of $W_{\mu \nu}^\Lambda(q,p,s)$ is found to be
\be
W_{\mu \nu}^\Lambda (q,p,s)= {1 \over 4 \pi} \int d^4 \xi e^{i q.\xi}
<0\vert J_\mu(0) \vert \Lambda(p,s) X><\Lambda(p,s) X\vert J_\nu(\xi)\vert 0>
\label{hadten}
\ee
where $J_\mu(\xi)$ is the electromagnetic current, $X$ is unobserved hadrons
(where the summation over all $X$ is assumed),
$q$ is the virtual photon momentum given by $q=q_1+q_2$, 
$p$ and $s$ are the momentum and spin of the polarised $\Lambda$ detected.

The polarised $\Lambda$ production cross section gets contribution
only from antisymmetric parts of leptonic and photonic tensor.
The antisymmetric part of the leptonic tensor is found to be $\cite {LU}$
\be
\tilde L_{\mu \nu}(q_1,q_2,s_1)= -2 i e^2 s_1 
\epsilon_{\mu\nu\alpha\beta} q_1^\alpha 
q_2^\beta
\label{lepten}
\ee
where again $q_1,q_2$ are the momenta of the incoming leptons and
$s_1$ is the spin of the polarised lepton(electron).
The photonic fragmentation tensor is not calculable
in Perturbative QCD(PQCD) as we do not know how to compute 
the matrix element of $em$(electro magnetic) current 
between $\Lambda$ states and the vacuum.  But this can be
parametrised using Lorentz covariance, gauge invariance, Hermiticity,
and parity invariance.  Hence the antisymmetric part 
of this photon fragmentation
tensor takes the following form $\cite {LU}$:
\be
\tilde W_{\mu \nu}^\Lambda(q,p,s)= {i\over p.q} 
\epsilon_{\mu \nu \lambda \sigma} 
q^\lambda s^\sigma \hat g_1^\Lambda(x,Q^2) + {i\over p.q} \epsilon_{\mu \nu 
\lambda \sigma} q^\lambda \left (s^\sigma- {s.q \over p.q} 
p^\sigma \right) \hat g_2^\Lambda(x,Q^2)
\label{hadexp}
\ee
where $x=2 p.q/Q^2$, $Q^2=q^2$ and $s^2=-1$.
Here the polarised fragmentation functions $\hat g_i^\Lambda(x,Q^2)$ 
are real and Lorentz 
invariant, hence they are functions of $x$ and $Q^2$.  

We consider the following asymmetric cross section such that only 
the $\hat g_1^\Lambda(x,Q^2)$ structure function is projected out:  
\be
\partud=\alpha^2 {\pi \over Q^2} x \hat g_1^\Lambda(x,Q^2) Cos(\theta)
\label{asym}
\ee   
Where $\alpha= e^2/4 \pi$, $\theta$ is the angle between 
produced hadron and the incoming electron. 
Here, $\uparrow \uparrow$ means that both incoming electron
and the produced hadron are paralelly polarised and $\uparrow \downarrow$
means that they are polarised anti-paralelly.  From the above
asymmetry it is clear that 
the polarised fragmentation function $\hat g_1^\Lambda(x,Q^2)$ can be 
measured only 
through angular distribution of polarised $\Lambda$ $\cite {BUR}$.  

Recall that in the DIS, the hadronic tensor appearing in the cross section
formula is factorised into
a perturbatively calculable hard part and non-perturbative matrix 
elements of quark and gluon operators sandwiched between hadronic 
states(called soft part) $\cite {QCD}$.  This factorisation ensures that 
the hard part contains no IR singularities.  Following similar 
factorisation procedure, one can write the photonic fragmentation 
tensor in terms of completely calculable hard part denoted 
by $H_{\mu\nu}^a(y,Q^2)$ and non-perturbative
operator matrix elements denoted by $D_a^\Lambda(z,Q^2)$ whose operator
definitions will be given later.  Hence, the factorisation formula
for photon fragmentation function reads as
\be
\tilde W_{\mu \nu}^{\Lambda(s)}(x,Q^2)= \sum_{a(h)} \int_x^1 {dy \over y} 
{\cal H}_{\mu \nu}^{a(h)} (x/y,Q^2)
\Delta D_{a(h)}^{\Lambda(s)}(y,Q^2)
\label{fact}
\ee
%where 
%\bea
%\Delta D_a^\Lambda(z,Q^2)&=&D_{a(\uparrow)}^{\Lambda(\uparrow)} (z,Q^2)
%-D_{a(\downarrow)}^{\Lambda(\uparrow)}(z,Q^2)
%\nonumber
%\eea
where $a$ runs over all the partons such as quarks, anti-quarks and gluons,
$h$ is the helicity of the parton.
$H_{\mu \nu}^{a(h)}(z,Q^2)$ is the hard part of the parton 
differential cross section for the
production of parton of type $a$ with polarisation $h$ and energy fraction
$z=2 p.q/Q^2$.
The soft part $\Delta D_{a(h)}^{\Lambda(s)}(y,Q^2)$ 
is the usual polarised parton 
fragmentation function whose operator definition and the physical
interpretation in terms of the partons will be discussed in the following. 
The above factorised formula ensures that 
the $H_{\mu \nu}^a(z,Q^2)$ is free of any
Infrared(IR) singularities.  In other words, if $H_{\mu\nu}^a(z,Q^2)$ is
free of any IR singularities, the tensor $W_{\mu \nu}^\Lambda(x,Q^2)$ 
is said to
be factorisable.  In the following we shall show that such
a factorisation exists and that it works to order $\alpha_s$
in the gluonic sector.  This is the by product of our analysis
on the evaluation of first moment of polarised 
gluonic contribution to polarised $\Lambda$ production.

Let us first work out the operator definition for $D_q^\Lambda(x,Q^2)$ to
leading order.
As we know the photon fragmentation tensor $W_{\mu \nu}^\Lambda(x,Q^2)$ is
not computable in the perturbation theory due to the non-perturbative
nature of hadrons(here $\Lambda$).  The physical 
interpretation of the photon fragmentation 
tensor
in terms of the parton fragmentation functions 
can be easily got if we work on the
light cone $\cite {JAFFE}$ since we are interested
in the limit $Q^2 \rightarrow \infty$(see below).  To make the 
computation simple 
and the physics more transparent,
let us assume that the currents appearing
in the photon fragmentation tensor are free field currents so that
formal manipulations can be done with out explicitly evaluating them
between the $\Lambda$ states and the vacuum.  
Note that dominant contribution to
this integral in $W_{\mu \nu}^\Lambda(x,Q^2)$ (eqn.(\ref {hadten})) 
in the limit $Q^2 \rightarrow \infty$ 
comes from the light cone region $\xi^2 \rightarrow 0$,
otherwise the exponential factor will kill the integral in 
this limit.  In this lightcone
limit $\xi^2 \rightarrow 0$, let us assume that only one of the field operators
(say quark) of each current is responsible for the fragmentation whatso 
ever be the
mechanism.   Hence the rest of the operator matrix elements
can be computed between free parton (say anti-quark) states 
to the leading order. 
After a little algebra we find that the quark contribution 
to the structure function 
$g_1^\Lambda(x,Q^2)$ to leading order on the light cone turns out to be
\be
g_1^\Lambda(x,Q^2)= 3{1 \over x}\sum_q e_q^2  \Delta D_q^{0\Lambda}(x,Q^2)
\ee
where
\be
\Delta D_q^{0\Lambda}(x,Q^2) ={x \over 8 \pi} \int d 
\xi^-  e^{-i \xi^- p^+/x}
< 0\vert \gamma_+ \gamma_5 \psi(0) \vert \Lambda X > < \Lambda X\vert
\bar \psi(\xi^-) \vert 0>
\label{firq}
\ee
where average over colour is implicit in the eqn.(\ref {firq}).
Similarly one can find out the anti-quark contribution to $g_1^\Lambda(x,Q^2)$.
Following the procedure adopted in reference $\cite {MAN1}$, we find that
the $\Delta D_q^{0\Lambda}(x,Q^2)$ measures the number of up($\uparrow$)
polarised
quarks fragmenting into up($\uparrow$) polarised $\Lambda$ minus 
the number of down($\downarrow$) polarised quarks fragmenting 
into up($\uparrow$) polarised $\Lambda$.   

Generalising the above
idea, one can express polarised quark, anti-quark and gluon fragmentation
functions in terms of quark and gluon operators as follows $\cite {MAN1},
\cite {JI}$:
\bea
\Delta D_{q+\bar q}^\Lambda(x,Q^2)&=& {x \over 8 \pi}\int e^{-i  \xi^- p^+ /x}
\left (< 0\vert \gamma_+ \gamma_5 \psi(0) \vert \Lambda X >
< \Lambda X\vert \bar \psi(\xi^-) \vert 0>
\right. \nonumber \\
&& \left .+< 0\vert \gamma_+ \gamma_5 \psi(\xi^-) \vert 
\Lambda X > < \Lambda X\vert \bar \psi(0) \vert 0> \right)
\label{matq}
\eea
\be
\Delta D_g^\Lambda(x,Q^2)\!=\! -{x^2 \over 4 \pi p^+}\int  e^{-i  \xi^- p^+ /x}
%\left (
< 0\vert \tilde G_a^{+\alpha}(0) \vert \Lambda X >
< \Lambda X\vert G^+_{a~\alpha}(\xi^-) \vert 0> 
%\right. \nonumber \\
%&& \left.-< 0\vert \tilde F_\alpha^{+a}(\xi^-) \vert \Lambda X >
%< \Lambda X\vert F_a^{\alpha+}(0) \vert 0> \right)
\label{matg}
\ee
where, sum over $X$ is assumed and $G_{\mu \nu}^a$ the usual field
strength tensor.  Also, averge over colour is implicit.  
We have also dropped the Wilson link operator as it becomes identity
in the lightcone gauge $A^+=0$ which we choose to work with.
The $\Lambda$ carries the momentum $p$ and spin $s$.
It is clear from the above operator definitions of the parton fragmentation
functions that the matrix elements of these
operators for hadronic states are not calculable.  But,
interestingly, they are calculable for the partonic states such as
quarks, anti-quarks and gluons in the perturbation theory as
one knows how these partonic operators act on the partonic states.
This fact that they are calculable for the partonic states is exploited
to compute the hard part $(H_{\mu \nu}^a(z,Q^2))$ of the factorisation formula.

First we compute the leading order contribution coming from the
operators to the polarised fragmentation function $\hat g_1^\Lambda(x,Q^2)$.
Using the factorisation formula and the normalisation that
$\Delta D_q^q(x,Q^2)=\delta(1-x)$ to lowest order,  we find that
the quark and the anti-quark contributions to $\hat g_1(x,Q^2)$ come 
form Fig. 1. which we denote by $\hat g_1^0(x,Q^2)$,  
\be
\hat g_{1q}^{0\Lambda}(x,Q^2)=3{1\over x} \sum_q e_q^2 \left( 
\Delta D_q^\Lambda(x,Q^2) + \Delta D_{\bar q}^\Lambda(x,Q^2)
\right)
\ee

From the above equation, it is clear that the polarised fragmentation
function $g_1^\Lambda(x,Q^2)$ to leading order $\alpha_s(Q^2)^0$ measures 
the polarised quark and anti-quark
fragmentation functions weighted by appropriate charge square 
factors($e_q^2$).

The $\alpha_s(Q^2)$ corrections to $g_1^\Lambda(x,Q^2)$ come from two sources: 
$\bf 1$. gluon bremstalung and virtual corrections to polarised quark and
anti-quark fragmenting into polarised $\Lambda$,  $\bf 2$.  
the polarised gluon fragmenting into polarised $\Lambda$.  
Here, we are interested only in the evaluation of polarised gluonic
contribution to the production.
Using the factorisation formula(eqn.(\ref {fact}), the 
gluonic contribution to $g_1^\Lambda(x,Q^2)$ 
can be formally written as 
\be
\hat g_{1g}^\Lambda(x,Q^2)={1\over x}\sum_q e_q^2 \int_x^1 {dy \over y} 
{\cal H}_g(x/y,Q^2) \Delta D_g^\Lambda(y,Q^2) 
\label{g1g}
\ee
We again use the factorisation formula to compute $H_{\mu \nu}^g(z,Q^2)$.
$H_{\mu \nu}^g(z,Q^2)$ gets contribution from order $\alpha_s(Q^2)$
onwards.  To compute this, we replace polarised $\Lambda$ by polarised gluon
in the eqn. (\ref {fact}).  With this replacement we find that 
the left hand side of the eqn. (\ref {fact}) is just 
the cross section for photon decaying into polarised 
gluon and a quark-antiquark pair. On the other hand, the right side 
involves the evaluation of quark operator for polarised gluonic states 
to order $\alpha_s(Q^2)$ and $g_{1q}^q(z,Q^2)$ and $\Delta D_g^g(z,Q^2)$
to lowest order $(\alpha_s(Q^2)^0)$.  Hence,
\be
{\cal H}_g(z,Q^2)= W_g(z,Q^2)-3 \Delta D_q^g(z,Q^2)
\label{glcf}
\ee

Let us now compute $W_g(z,Q^2)$ appearing in the eqn.(\ref {glcf}).
$W_g(z,Q^2)$ can be computed from the polarised
cross section for $e^+ e^- $ annihilating into the polarised gluon:  
\be
\partg={i s \over 4 q_1.q_2} \tilde L_{\mu \nu} (q_1,q_2,s_1) {1 \over Q^4}
\epsilon_{\mu\nu\lambda\sigma} q^\lambda p^\sigma_g 
{Q \over p_g.q} H_g(p_g,q)
\label{asyg1}
\ee
where, $z=2p_g.q/Q^2$, 
$p_g$ is the momentum of the polarised gluon and  
$\theta$ is the angle between produced
gluon and the incoming electron.
$H_g(p_g,q)$ appearing in the 
above equation (eqn. (\ref {asyg1})) is 
found to be
\be
H_g(p_g,q)= {Q \over 32 (2 \pi)^3} \int dx_2 {\cal P}_g^{\mu \nu} \vert M^g 
\vert^{2}_{\mu \nu}
\label{hadg}
\ee
where $x_2=2p_q.q/Q^2$ and the projector ${\cal P}_g^{\mu \nu} =i 
\epsilon_{\mu \nu \lambda \sigma} p_g^\lambda q^\sigma/2 p_g.q$.
The projected matrix element square ${\cal P}_g.\vert M^g\vert^2$ 
is computed from 
the Fig. 2(with gluon polarised) and is given by
\bea
{\cal P}_g^{\mu \nu} \vert M^g \vert^{2}_{\mu \nu}\!\!&=&\!\!3 C_2(R){4 (2 \pi)^3
e_q^2 \alpha \alpha_s \over \pi}\!\! \left({m_g^2 Q^2 -st 
\over (s+t)^2}\right)\! \!
\left[ 2{(s+t)(s+t-m_g^2-Q^2) \over s t} \right. \nonumber \\
&& \left. + {4 m_g^2 Q^2 - 2 m_g^2 s - 2 Q^2 s +s^2 -t^2 \over t^2} 
\right. \nonumber \\
&& \left. + {4 m_g^2 Q^2 - 2 m_g^2 t - 2 Q^2 t +t^2 -s^2 \over s^2} 
 \right]
\label{prhadg}
\eea
where the Mandelstam variables $s=(p_{\bar q}+p_g)^2$,$t=(p_q+p_g)^2$ and
$u=(p_q+p_{\bar q})^2$ satisfy $s+t+u=m_g^2+Q^2$.  Also note that
$s=Q^2(1-x_2)$.  Substituting the eqns. (\ref {prhadg}) and (\ref {hadg}) 
in the eqn. (\ref {asyg1}), 
we find that the
asymmetry in the polarised gluon emission turns out to be
\be
\partG={3 C_2(R) e_q^2 \alpha^2 \alpha_s \over  Q^2} \left[ (2-z) 
\log\left({1+ \beta_z\over 1- \beta_z} 
\right) - 2 (2 -z) \beta_z \right] Cos(\theta)
\label{asyg2}
\ee
where $z= 2 p_g.q/Q^2$, $\beta_z=(1- 4 m_g^2/Q^2 z^2)^{1/2}$ and 
$\theta$ here is the angle between the
outgoing polarised gluon and incoming lepton(electron).
The above result shows that there are no soft singularities.
The small nonzero gluon mass is used to regulate collinear
divergence.  Since we are looking at the polarised gluon production,
there is no virtual correction to this order, hence
there is no UV divergences.   
From the eqn. (\ref {asyg2}), we find that
\be
W_g(z,Q^2)=3 C_2(R){\alpha_s \over 2 \pi} 2\left[  (2-z) 
\log({Q^2 z^2 \over m_g^2 }) - 2 (2-z)\right]
\label{cg}
\ee

Now, let us compute the matrix element $D_q^g(z,Q^2)$ appearing in the
eqn. (\ref {glcf}). We can compute this using the eqn. (\ref {matq}) 
with $\Lambda$ 
replaced by polarised gluon.
The diagrams contributing to this matrix element are given in Fig.3.	
This is found to be
\bea
\Delta D_{q+\bar q}^g(z,Q^2)&=& 4  C_2(R) \alpha_s z~ \int 
{d^d k \over (2 \pi)^d} 2 \pi \delta(k^2) 
2 \pi \delta( k^+ -  p_g^+ + p_g^+/z)\\ \nonumber 
&& Tr\left[ \gamma^+ \gamma_5 { \not k -\not p_g  \over (k-p_g)^2 -i\epsilon} 
\not \epsilon^* \not k \not \epsilon 
{\not k-\not p_g \over (k-p_g)^2 +i \epsilon} \right]
\label{triangle1}
\eea
The above expression is UV singular and we regulate this
divergence in Dimensional Regularisation method.  A small gluon mass
$m_g$ is introduced to regulate IR divergence.  
From the detla functions appearing in the above equation we find that 
$k^+=p_g^+(z-1)/z$ and $k^-=k_\perp^2 z /(2 p_g^+(z-1))$. 
Using these constraints, the $k^+$ and $k^-$ integrations can be 
safely done.  The $k_\perp^2$ integration can 
also be done and the result is
\be
\Delta D_{q+\bar q}^g(z,Q^2) =  C_2(R){\alpha_s \over 2 \pi} 2\left[
(2-z) \log \left ( {\mu_R^2 z^2 \over mg^2 (1-z)}\right )
-z \right] 
\label{triangle2}
\ee
The UV divergence appearing in the eqn.(\ref{triangle1}) is
renormalised in $\bar {MS}$ scheme, hence the appearance of substration  
scale $\mu_R^2$.

Substituting the eqn.(\ref {triangle2}) in eqn. (\ref {glcf}), 
we find that the gluonic coefficient function is
\be
{\cal H}^g(z,Q^2) = 3C_2(R){\alpha_s \over 2 \pi}2 \left[
(2-z) \log \left ( {Q^2 \over \mu_R^2}\right )
+ (2-z) \log \left ( 1-z \right) +3z -4 \right] 
\ee
where $\mu_R$ is the scale at which the gluon fragmentation
operator matrix element is renormalised.
From the above equation it is clear that the collinear
divergence appearing in the cross section formula
cancels against that appearing in the matrix element
leaving the hard part $H_g(z,Q^2)$ free of any $IR$ divergences.
This is the proof of factorisation in the gluonic sector to
order $\alpha_s(Q^2)$.  Substituting the above equation in
eqn.(\ref {g1g}), we get the gluonic contribution to
$g_1^\Lambda(x,Q^2)$ and hence the gluonic contribution to
the asymmetry given in the eqn. (\ref {asym}).

In this paper we have systematically analysed the importance of
gluons to the production of polarised $\Lambda$ in an inclusive
$e^+ e^-$ annihilation.  We have done this using
the AP equation for the gluon fragmentation function.
We have computed the gluonic contribution to order $\alpha_s$
to the polarised fragmentation function $g_1^\Lambda(x,Q^2)$
appearing in the asymmetry of the polarised cross section
using the factorisation formula.  The factorisation that
the cross section can be factored into a hard and a soft part
has been demonstrated in the gluonic sector.

It is a pleasure to thank Prof M.V.N. Murthy for his 
constant encouragement and fruitful discussions.
I would like to thank Prof H.S. Mani for his invitation 
to Metha Research Institute where this work was formulated
and I am also thankful to Prof H.S. Mani, Prof R. Ramachandran 
and Prof K. Sridhar for their carefull reading of the 
manuscript and valuable comments.
I thank Dr. Prakash Mathews for his valuable discussions.

\newpage
{\bf \Large Figure Captions:}
\begin{enumerate}
\item
Graph contributing to $e^-(q_1) e^+ (q_2)\rightarrow \bar q(p_{\bar q})
q(p_q)$.
\item
Graphs contributing to $e^-(q_1) e^+ (q_2)\rightarrow \bar q(p_{\bar q}) 
q(p_q)
g(p_g)$.
\item
Graph contributing to $\Delta D_q^g(x,Q^2)$
\end{enumerate}
\end{document}